\RequirePackage{fix-cm}
\documentclass[smallextended]{svjour3}       
\usepackage{graphicx}

\usepackage{mathrsfs}

\usepackage{amsmath}
\usepackage{amssymb}
\usepackage{placeins}
\usepackage[]{qcircuit}

\newcommand{\ket}[1]{|#1\rangle}
\graphicspath{{./pics/}}

\usepackage{caption}
\usepackage{subcaption}
\usepackage[english]{babel}

\begin{document}

\title{Nondestructive classification of quantum states using an algorithmic quantum computer}


\titlerunning{Nondestructive classification of quantum states}        

\author{D. V. Babukhin \and
A.\,A.~Zhukov         \and
     W.\,V.~Pogosov
}


\institute{D.\,V.~Babukhin \at
  Dukhov Research Institute of Automatics (VNIIA), 127055 Moscow, Russia \\
    Russian Quantum Center (RQC), 143026 Moscow, Russia\\
    Lomonosov Moscow State University (MSU), 119991 Moscow, Russia \\
  \and
A.\,A.~Zhukov \at
              Dukhov Research Institute of Automatics (VNIIA), 127055 Moscow, Russia \\
              National Research Nuclear University (MEPhI), 115409 Moscow, Russia
\and
      W.\,V.~Pogosov   \at
    Dukhov Research Institute of Automatics (VNIIA), 127055 Moscow, Russia \\
    Institute for Theoretical and Applied Electrodynamics, Russian Academy of
    Sciences, 125412 Moscow, Russia\\
    Tel.: +7(926)359-6034\\
    Fax.: +7(499)978-0903\\
    \email{Walter.Pogosov@gmail.com}           
  }

\date{Received: date / Accepted: date}

\maketitle

\begin{abstract}
Methods of processing quantum data become more important as quantum computing devices improve their quality towards fault tolerant universal quantum computers. These methods include discrimination and filtering of quantum states given as an input to the device that may find numerous applications in quantum information technologies. In the present paper, we address a scheme of a classification of input states, which is nondestructive and deterministic for certain inputs, while probabilistic, in general case. This can be achieved by incorporating phase estimation algorithm into the hybrid quantum-classical computation scheme, where quantum block is trained classically. We perform proof-of-principle implementation of this idea using superconducting quantum processor of IBM Quantum Experience. Another aspect we are interested in is a mitigation of errors occurring due to the quantum device imperfections. We apply a series of heuristic tricks at the stage of classical postprocessing in order to improve raw experimental data and to recognize patterns in them. These ideas may find applications in other realization of hybrid quantum-classical computations with noisy quantum machines.


\end{abstract}

\section{Introduction}

Machine learning is a computing paradigm, where recognition of patterns in available data plays a central role, but the computing system is not explicitly programmed; many examples indeed demonstrate success of this approach to real-world problems. Quantum machine learning is an emergent technology based on the assumption that quantum resources can be useful in the pattern analysis, see, e.g., Refs. \cite{Lloyd1,Schuld1,Biamonte,Amin,Preskill,Adcock}. Quantum algorithms within such applications can be used as a part of a larger computation scheme which also incorporates classical blocks.

There are two major approaches for the construction of a quantum block in such schemes -- it can be represented either by quantum annealer or by algorithmic quantum computer \cite{Biamonte}. Most of the proposals unfortunately are characterized by input/output bottlenecks occurring at stages of encoding classical data into quantum states and decoding them back  \cite{Aaronson,Mosca}. However, these bottlenecks seem to be not severe in the case input states are quantum \cite{Wiebe1,Wiebe2}. The role of the quantum machine is to recognize their underlying patterns, which may have no classical counterpart (for example, characteristics of quantum entanglement), and then to classify these states or filter them. Let us stress that the classification of quantum states is supposed to play a crucial role in quantum metrology and sensing \cite{sensing}. For instance, in quantum illumination problem one has to operate with the entangled photonic states and to reveal their characteristics \cite{Illum1,Illum2}. Another possible source of quantum data can be a quantum simulator or another quantum computer (for example, more noisy and/or of a larger size) \cite{Biamonte}.

Machine learning tasks can be roughly divided into supervised and unsupervised. In the present paper we address a hybrid quantum-classical approach to the problem of classification of input quantum states, where quantum block is trained classically with the set of labeled input vectors (supervised learning). An essential ingredient of the model we consider is a phase estimation algorithm embedded into the quantum part of the computational scheme. Using ancilla qubits it is possible to extract information about quantum state without doing a direct measurement of the qubits encoding this state. It is thus possible to make a classification of certain input quantum states both nondestructively and deterministically. For general input states, the classification is probabilistic. This idea is motivated by the recent suggestion on simulation of perceptron on a quantum computer \cite{PetruccioneNN}.

We also perform proof-of-principle realization of our scheme with real superconducting quantum computer of IBM Quantum Experience available through the cloud service. Its performance, as well as performances of existing quantum computers based on other physical realizations, is limited by imperfections of quantum hardware, which include effects of decoherence and quantum gate errors. This limitation restricts possible realizations of quantum machine learning algorithms to few-qubit examples, see, e.g., Refs. \cite{photon,nmr}. We therefore address a rather simple toy model, which is associated with the classification of maximally entangled two-qubit states. In order to obtain a valuable information from raw experimental data affected by noise, we apply a series of tricks based on classical posprocessing which are also associated with pattern recognition. These ideas can be of interest in a general context of hybrid quantum-classical computation, which attracts a lot of attention now, see, e.g., Refs. \cite{QAOA,theory,Peruzzo,Preskill,variat,ML}.

This paper is organized as follows. In Section II we explain basic ideas behind the approach used. In Section III we present an explicit treatment of a toy model dealing with the classification of two-qubit maximally entangled states. In Section IV we describe the realization of this toy model on superconducting quantum computer of IBM Quantum Experience and apply different approaches to mitigate the effect of errors. We conclude in Section V.

\section{Phase estimation algorithm in classification problems}

Programmable quantum computers operate with data encoded into quantum states. An example of the potential applications for quantum computers is a classification of states given as input, according to some criterion or criteria. In order to accomplish this task one has to construct a circuit which signals out if a state belongs to one of predefined classes.
Another example is associated with filtering problem -- quantum device must nondestructively pass a state which belongs to a predefined class and should also signal this event out.
The problem how to construct such a circuit is obvious only for trivial cases and it is not simple for more complex quantum states.

One of the possible solutions is to use ideas from the machine learning field. For example, it is reasonable to construct a quantum circuit with some limited number of free parameters which enter certain blocks of the algorithm. Then, the quantum algorithm can be "trained" by sending training states to the input, tuning the parameters and finding their optimal values allowing for the desirable classification, which can include multiple groups.
It is difficult to implement the training as a purely quantum procedure, so that this part of the whole scheme might be accomplished classically, i.e., through the classical computer. The scheme, in this case, represents one of the numerous examples of a hybrid quantum-classical computations.
The classical training procedure can be based on various methods, such as grid search, Monte Carlo method, or gradient descent method.

\begin{figure}[h!]
	\[
	\Large 
	\Qcircuit @C=1.0em @R=.6em 
	{
		&\lstick{\ket{0}} & \gate{H} & \qw & \qw & \ctrl{3} & \qw & \multigate{1}{QFT^{-1}} & \qw & \meter \\ 
		&\lstick{\ket{0}} & \gate{H} & \qw & \ctrl{2} & \qw & \qw & \ghost{QFT^{-1}} & \qw & \meter \\
		... \\
		&\lstick{\ket{\psi}} & \qw {/} & \qw & \gate{U(\vec{\omega})} & \gate{U(\vec{\omega})^{2}} & \qw & \lstick{\cdots} & \qw & \\ 
	}
	\]
	\caption{A schematic view of the quantum circuit. $\ket{\psi_{in}}$ is an input state, $U(\vec{\omega})$ is a parametrized unitary, where $\vec{\omega}$ is a set of tunable parameters to be adjusted during the training procedure.}\label{circuit}
\end{figure}
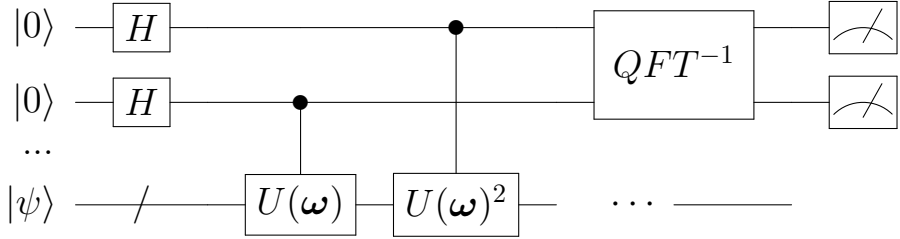

In the present paper, for the quantum part of the scheme we adopt ideas based on the phase estimation algorithm, which enables to get information about an input state without doing a direct measurement of the qubits encoding this state, but instead exploits ancilla qubits. For certain input states, the classification can be made both nondestructive and  deterministic. The quantum block of this circuit is shown schematically in Fig. \ref{circuit}, where $U(\vec{\omega})$ is a unitary operator parametrized by a set of tunable parameters $\vec{\omega}$ to be adjusted during the training procedure. If the input state is an eigenstate of $U(\vec{\omega})$, the measurements of ancilla qubits do not destroy it, so $\ket{\psi}$ is passed nondestrucvitely through the scheme (apart of a general phase it obtains). Moreover, in this case, the measurements of ancilla qubits are deterministic, provided the eigenvalue of $\ket{\psi}$ is $\exp(2i \pi n/2^{N_a})$, where $N_a$ is the number of ancilla qubits, whereas $n$ is an integer number ranging from 0 to $2^{N_{a}} - 1$. The inverse statement is also true: deterministic results of ancilla's measurement is possible only if the input state is one of the eigenstates of $U(\vec{\omega})$ and its eigenvalue is of the form $\exp(2i \pi n/2^{N_a})$.

Hence, if there are two input states each being eigenstates of $U(\vec{\omega})$ with different eigenvalues of the above type, it is possible to classify these states both nondestructively and deterministically by doing measurements of ancillas. Otherwise the classification is probabilistic: the probability to get a set of 0 and 1 corresponding to the eigenstate of $U(\vec{\omega})$ with given eigenvalue $\exp(2i \pi n/2^{N_a})$ is  the sum of overlaps between $\ket{\psi}$ and all mutually orthogonal eigenstates of $U(\vec{\omega})$ characterized by this particular eigenvalue. If $\ket{\psi}$ is the eigenstate of $U(\vec{\omega})$, the classification is nondestructive, but probabilistic, in general case. Notice that the nondestructive character of state transfer through the circuit can be probed by the SWAP test.

We now discuss the same problem, but from another perspective. Let us assume that we have $M$ orthogonal input states. We may try to perform an ideal classification of these states, i.e., to construct an operator $U(\vec{\omega})$, for which these states are eigenstates and, moreover, the results of ancilla's measurements allow for the unambiguous deterministic discrimination between them. Let us stress that such a circuit provides a nondestructive and deterministic classification among given set of $M$ input states, while a general input state is classified probabilistically. In the latter case, through the repeating measurements, we may recognize which of the $M$ states of the training set the input state is closer to. It is clear that the minimum necessary number of ancilla qubits is determined by the condition $2^{N_a}\geqslant M$. Apparently, requirements for the operator $U(\vec{\omega})$ for such a classification are quite restrictive. Alternatively, it is possible to find such a $U(\vec{\omega})$, which yields a nondestructive but probabilistic classification of $M$ orthogonal training states. Again, the nondestructive character of the input state transfer through the circuit can be verified by SWAP test.

The problem of efficient construction of desirable $U(\vec{\omega})$ is far from being obvious. In principle, it is possible to try a brute-force strategy, which seems rather universal: one may use a fixed entangler of all qubits of the register and to apply it multiple times, but to insert a set of single-qubit rotations between each application of the entangler; rotation angles can be treated as variational parameters. A similar approach was utilized in Ref. \cite{variat} for the preparation of variational many-body states for the modelling of molecules. It is then possible to optimize some error function in order to minimize a level of "destructiveness" or "non-determinism" of the classification. Another possible strategy is to rely on heuristics when finding suitable form of $U(\vec{\omega})$, which depends on characteristics of vectors from the training set. Below we discuss a toy model, which contains all essential ingredients of the scheme we discuss and can be tested with existing quantum machines. Within this simple example, we follow the heuristic approach for the construction of a proper operator $U(\vec{\omega})$.

\section{Toy model: classification of maximally-entangled two-qubit states}

Let us consider four possible input states defined as two-qubit maximally entangled states. In other words, we assume that there are four training vectors, which are Bell states $\ket{\Phi_{\pm}}$ and $\ket{\Psi_{\pm}}$, given by
\begin{eqnarray}
\ket{\Phi_{\pm}} = \frac{1}{\sqrt{2}}(\ket{00} \pm \ket{11}),\notag \\
\ket{\Psi_{\pm}} = \frac{1}{\sqrt{2}}(\ket{10} \pm \ket{01}).
\label{Bell}
\end{eqnarray}
Our aim is to construct an ideal classification scheme allowing for the nondestructive and deterministic classification of these four states into two classes $\ket{\Phi_{\pm}}$ and $\ket{\Psi_{\pm}}$.

The states of these two classes differ from each other by their "internal structure" reflected in the probabilities to be in the orthogonal states of computational basis, which is not sensitive to the phases. Therefore, it is perspective to construct $U$ on the basis of rotations around $z$ axis. We thus parametrize $U$ as $U=U_{z1}(\omega_1)U_{z2}(\omega_2)$, where indices 1 and 2 refer to the qubit number and $U_{z}(\omega)=\begin{bmatrix}
e^{-i\pi\omega/2} & 0 \\
0 & e^{i\pi\omega/2} \\
\end{bmatrix}$ is a single-qubit rotation around $z$ axis.

We first show explicitly that such a parametrization for $U$ gives a desirable result and also determine optimal values of $\omega_1$ and $\omega_2$ yielding nondestructive and deterministic classification. We then do the same work using the real quantum computer by finding such optimal parameters through the grid search that can be treated as a learning procedure.

\begin{figure}[h!]
	\center
	\[
	\Qcircuit @C=1em @R=.7em
	{
		\lstick{\ket{0}}	& \gate{H} & \qw & \qw & \qw & \ctrl{1} & \ctrl{2} \qw & \gate{H} & \qw & \qw & \meter \\
		\lstick{} & \qw & \qw & \qw & \qw &  \gate{R_{z}(w_{1})} & \qw & \qw & \qw & \qw & \qw  \\
		\lstick{} & \qw & \qw & \qw & \qw & \qw & \gate{R_{z}(w_{2})} &\qw & \qw & \qw & \qw
		\inputgroupv{2}{3}{.8em}{.8em}{\ket{\psi}}\\
	}
	\]
	\caption{A quantum circuit for the case of two-qubit input states (see in the text).}
\label{2qcircuit}
\end{figure}
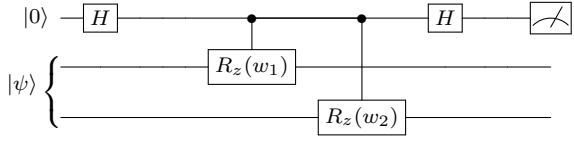

It is easy to see that $\ket{\Phi_{\pm}}$ are eigenstates of $U$ provided $\omega_1+\omega_2=2k$, where $k$ is an integer number. The eigevalue of $U$ for both $\ket{\Phi_{+}}$ and $\ket{\Phi_{-}}$ is the same, $U_{\Phi}=e^{-i\pi k}$. Similarly, $\ket{\Psi_{\pm}}$ are eigenstates of $U$ provided $\omega_1-\omega_2=2q$, where $q$ is an integer number; while the eigenvalue of $U$ for both $\ket{\Psi_{+}}$ and $\ket{\Psi_{-}}$ is the same, $U_{\Psi}=e^{-i\pi q}$. Let us choose $p$ and $q$ in such a way as to make $U_{\Psi}$ and $U_{\Phi}$ different from each other, which is necessary for the classification to work. Obviously, parities of $p$ and $q$ must be opposite. We may choose, for instance, $k=0$ and $q=1$, which leads to $U_{\Phi}=-U_{\Psi}=1$, $\omega_1=-\omega_2=1$. Fortunately, for our simplistic toy model both eigenvalues we found fall automatically into the discrete set, which enables for a deterministic classification. This can be achieved using a single ancilla. The whole quantum scheme for this case if shown in Fig. \ref{2qcircuit}. For the input state $\ket{\Phi_{\pm}} \otimes \ket{0}$, the output state at the end of the circuit is $\ket{\Phi_{\pm}} \otimes \frac{1}{2} ((1+U_{\Phi})\ket{0}+(1-U_{\Phi})\ket{1}) = \ket{\Phi_{\pm}} \otimes \ket{0}$. For the input state $\ket{\Psi_{\pm}} \otimes \ket{0}$, the output is $\ket{\Psi_{\pm}} \otimes \frac{1}{2} ((1+U_{\Psi})\ket{0}+(1-U_{\Psi})\ket{1}) = \ket{\Psi_{\pm}} \otimes \ket{1}$. Thus, we see that indeed nondestructive and deterministic classification of two groups of input states is possible, since for $\ket{\Phi_{\pm}}$ the probability $P_0(\ket{\Phi_{\pm}})$ to find ancilla in the state $\ket{0}$ is exactly 1, while for $\ket{\Psi_{\pm}}$ the probability $P_0(\ket{\Psi_{\pm}})$ to find ancilla in the state $\ket{0}$ is exactly 0. The scheme basically performs a parity check and the parity is to be considered as a "quantum pattern".

For the input two-qubit state of a general form
\begin{equation}
\ket{\Psi} = \alpha\ket{00} + \beta\ket{01} + \gamma\ket{10} + \delta\ket{11}
\label{genpsi}
\end{equation}
after some straightforward calculations we obtain the expression for probability $P_0(\ket{\Psi})$ to find ancilla in the state $\ket{0}$ provided optimal $\omega_1$, $\omega_2=1$ are incorporated into the circuit
\begin{equation}
P_0(\ket{\Psi}) = \frac{1}{2} + \frac{|\alpha|^{2} + |\delta|^{2}}{2} - \frac{|\beta|^{2} + |\gamma|^{2}}{2}.
\label{P}
\end{equation}
It can be rewritten as
\begin{equation}
P_0(\ket{\Psi}) =\frac{1}{2}+\frac{1}{2}\left( |\langle \Phi_{+}|\Psi \rangle|^2 +
|\langle \Phi_{-}|\Psi \rangle|^2 - |\langle \Psi_{+}|\Psi
\rangle|^2-|\langle \Psi_{-}|\Psi \rangle|^2 \right).
\label{P0}
\end{equation}
In this general case, the scheme works as a probabilistic classifier, and the classification occurs according to the distance between the input state and two subspaces, in which $\ket{\Phi_{\pm}}$ and $\ket{\Psi_{\pm}}$ form local bases. We stress that $P_0(\ket{\Psi})$ is no longer exactly 0 or 1, while a measurement of the ancilla cannot be treated as nondestructive. A nondestructive classification is possible between quantum states of two classes, $\alpha\ket{00} + \delta\ket{11}$ and $\beta\ket{01} + \gamma\ket{10}$.

\begin{figure}[h!]
	\begin{subfigure}{0.5\linewidth}
		\center{\includegraphics[width=1.0\linewidth]{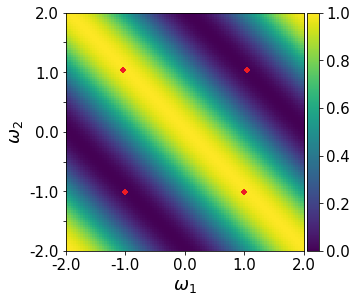}}
		\caption{}
		\label{fig: a}
	\end{subfigure}\hspace{0.0cm}
	\begin{subfigure}{0.5\linewidth}
		\center{\includegraphics[width=1.0\linewidth]{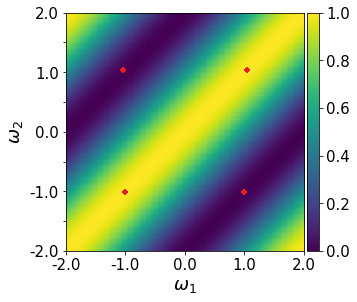}}
		\caption{}
		\label{fig: b}
	\end{subfigure}
	\caption{Probability patterns $P_0(\ket{\Psi}; \omega_1,\omega_2)$ for $\ket{\Psi}=\Phi_{\pm}$ (a) $\ket{\Psi}=\Psi_{\pm}$ (b). Points, where a discrimination between two pairs of Bell states is done nondestructively in a single measurement of the ancilla qubit, are marked by red.}
	\label{anatfig}
\end{figure}

Now let us come back to the previous stage and consider the learning procedure. If explicit treatment is impossible, optimal values of $\omega_1, \omega_2$ have to be determined from the results of measurements of ancillas. Let us introduce probability $P_0(\ket{\Psi}; \omega_1,\omega_2)$ to find the ancilla in the state $\ket{0}$ for general $(\omega_1, \omega_2)$ and for the input state $\ket{\Psi}$. This quantity is a generalization of $P_0(\ket{\Psi})$ given by Eq. (\ref{P}) and it can be written as
\begin{eqnarray}
P_0(\ket{\Psi}; \omega_1,\omega_2) = \frac{1}{2} + \frac{|\langle \Phi_{+}|\Psi \rangle|^2 +
|\langle \Phi_{-}|\Psi \rangle|^2}{2}\cos(\frac{\pi}{2}(\omega_{1} + \omega_{2})) + \notag \\
\frac{|\langle \Psi_{+}|\Psi
\rangle|^2 + |\langle \Psi_{-}|\Psi \rangle|^2}{2}\cos(\frac{\pi}{2}(\omega_{1} - \omega_{2})).
\label{Pcos}
\end{eqnarray}
The training procedure consists in finding optimal $(\omega_1, \omega_2)$ by evaluating both $P_0(\ket{\Phi_{\pm}}; \omega_1,\omega_2)$ and $P_0(\ket{\Psi_{\pm}}; \omega_1,\omega_2)$ and extracting points in the $(\omega_1, \omega_2)$ space, where the first quantity is exactly 1, while the second quantity is exactly 0 (or vice versa). Values of $(\omega_1, \omega_2)$ can be tuned by the classical computer, while quantum algorithm is implemented with the quantum computer. The brute-force method to determine optimal $(\omega_1, \omega_2)$ is a grid search. In the next Section we perform such a search using the real quantum computer. The experimental results will be compared with the explicit treatment. In order to facilitate this comparison, in Fig. \ref{anatfig}, we show the results of our calculations for $P_0(\ket{\Phi_{\pm}}; \omega_1,\omega_2)$ and $P_0(\ket{\Psi_{\pm}}; \omega_1,\omega_2)$ based on Eq. (\ref{Pcos}). From this figure we again see that there are values of $\omega_{1}$, $\omega_{2}$, supporting a discrimination between two pairs of Bell states in a single measurement.

\section{Implementation on a noisy quantum device}

\subsection{Quantum circuit}

Having a simple algorithm at hand, we perform proof-of-principle realization on a currently available quantum device. An additional important issue we are interested in is an error mitigation in hybrid quantum-classical computation schemes, so we consider the realization of a given algorithm as a playground for this quite general problem.

We use 16-qubit IBMqx5 superconducting quantum chip, which is available through the cloud service within the IBM Quantum Experience project. The realization of our scheme is illustrated in Figs. 4 and 5. Figure 4 shows the schematic image of the chip. The qubits utilized in our quantum algorithm are shown by the red color. The quantum circuit itself is presented in Fig. 5. Due to the limitations in connectivity, the quantum circuit includes an additional SWAP gate required to interchange quantum states of two physical qubits. Note that this gate is composed of three CNOT gates and it therefore provides an additional significant contribution to the total error rate.

\subsection{Raw data}

State-of-the-art quantum computers still suffer from decoherence problem, as well as imperfections of quantum gates and readouts. In order to use such devices for realization of quantum algorithms one has to deal with accumulation of errors.  It is worth discussing sources of errors for quantum circuits of different lengths under the realization on available superconducting quantum devices. Roughly, they can be divided into readout errors, quantum gate errors and a bare influence of decoherence, which are characterized as follows:

(i) Readout error is typically of the order of $10^{-2}$;

(ii) Average gate errors is of the order of $10^{-3}$. It is also known that errors of two-qubit gates are nearly one order of magnitude larger than that of single-qubit gates;

(iii) Longitudinal and transverse relaxation times of individual qubits are typically tens of microseconds. They must be compared to typical timescales of individual quantum gates. This time for single-qubit gates is nearly $80$ ns and the duration of two-qubit gates is about $300$ ns, there is also $10$ ns buffer between two gates.

To partially suppress or mitigate the errors, different tricks have been suggested \cite{mitig1,mitig2,mitig3,mitig4}. These tricks are usually efficient in the regime of low error rate, which is achieved provided shallow quantum circuits are used within the schemes of quantum-classical computation. In contrast, the implementation of our toy model is already associated with the quantum circuit which is not so shallow. Therefore, the error rates in our experiments are relatively high, while the dominant contribution is provided by CNOT errors. We therefore use a series of tricks based on classical postprocessing techniques applied for the output from a noisy quantum device. Since our final goal is to find experimentally the probability patterns of the form similar to the theoretical ones depicted in Fig. 3, in our treatment we also use certain analogies with a problem of image denoising. Thus, we again address the problem of pattern recognition, but now classically.


\begin{figure}[h!]
	
	\center{\includegraphics[width=1.0\linewidth]{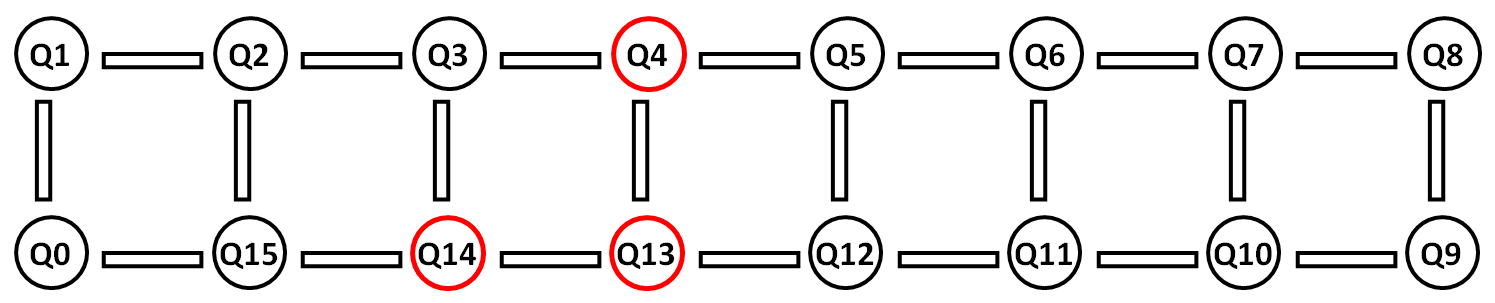}}
	\caption{Qubit connectivity map of the 16-qubit quantum chip IBMqx5. Qubits used for the scheme implementation are marked by the red color.}
	
\end{figure}

\begin{figure}[h!]
	\[
	\Qcircuit @C=1em @R=1.4em
	{
		\lstick{q4}	& \gate{H}  & \qw & \qw & \ctrl{2} & \qw & \ctrl{2} \qw & \gate{H} & \qw & \qw & \qw & \meter \\
		\lstick{q14} & \qw & \qw & \qw & \qw & \qswap & \qw & \qw &  \qw & \qw & \qw & \qw \\
		\lstick{q13} & \qw &\qw & \qw & \gate{R_{z}(w_{1})} & \qswap & \gate{R_{z}(w_{2})} & \qw & \qw & \qw & \qw & \qw
	}
	\]
	\caption{The quantum circuit implemented in real quantum processor IBMqx5. Pairs of symbols $\times$ denote SWAP operation on corresponding pair of qubits used to circumvent the limitations of connectivity of the chip.}
	
\end{figure}
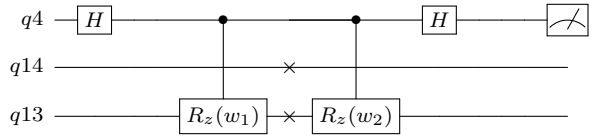

Figure 6 shows the results for $P_0(\ket{\Psi_{-}}; \omega_1,\omega_2)$ obtained from IBM classical simulator (left panel), which does not take into account device imperfections, and the real quantum machine (right panel). The results from the classical simulator are, of course, the same, within the computational accuracy and disregarding discretization, as the ones obtained analytically (see Fig. 3 b). Both experimental and theoretical maps contain $40 \times 40$ points. There were 8192 measurements for each point. We have chosen the state $\ket{\Psi_{-}}$ among the four possibilities in order to illustrate our results and ideas on error mitigation; the results for the remaining three states are rather similar. The comparison of the experimental and theoretical data shows that the agreement is not satisfactory -- the experimental data even for our toy classification model are heavily damaged by the noise.
Particularly, the experimental probabilities tend to approach $0.5$ instead of being distributed from 0 to 1.
Moreover, the experimental probability pattern also lacks "connecting bridges between islands": the exact pattern contains diagonal areas with high values of $P_0(\ket{\Psi_{-}}; \omega_1,\omega_2)$, while in the experimental data these diagonal areas are dissociated into five separate islands with suppressed values of $P_0(\ket{\Psi_{-}}; \omega_1,\omega_2)$ between them. Nevertheless, in the next subsections we are going to apply a combination of tricks in order to extract valuable information from so noisy raw data.

A poor quality of experimental data is the reason why we restricted ourselves to an oversimplified classification problem with few qubits only among 16 qubits of the device. Indeed, classification of quantum states involving larger number of qubits implies application of much larger number of two-qubit gates which provide the main contribution to the total error rate.

\begin{figure}[h!]
	\center{\includegraphics[width=1.0\linewidth]{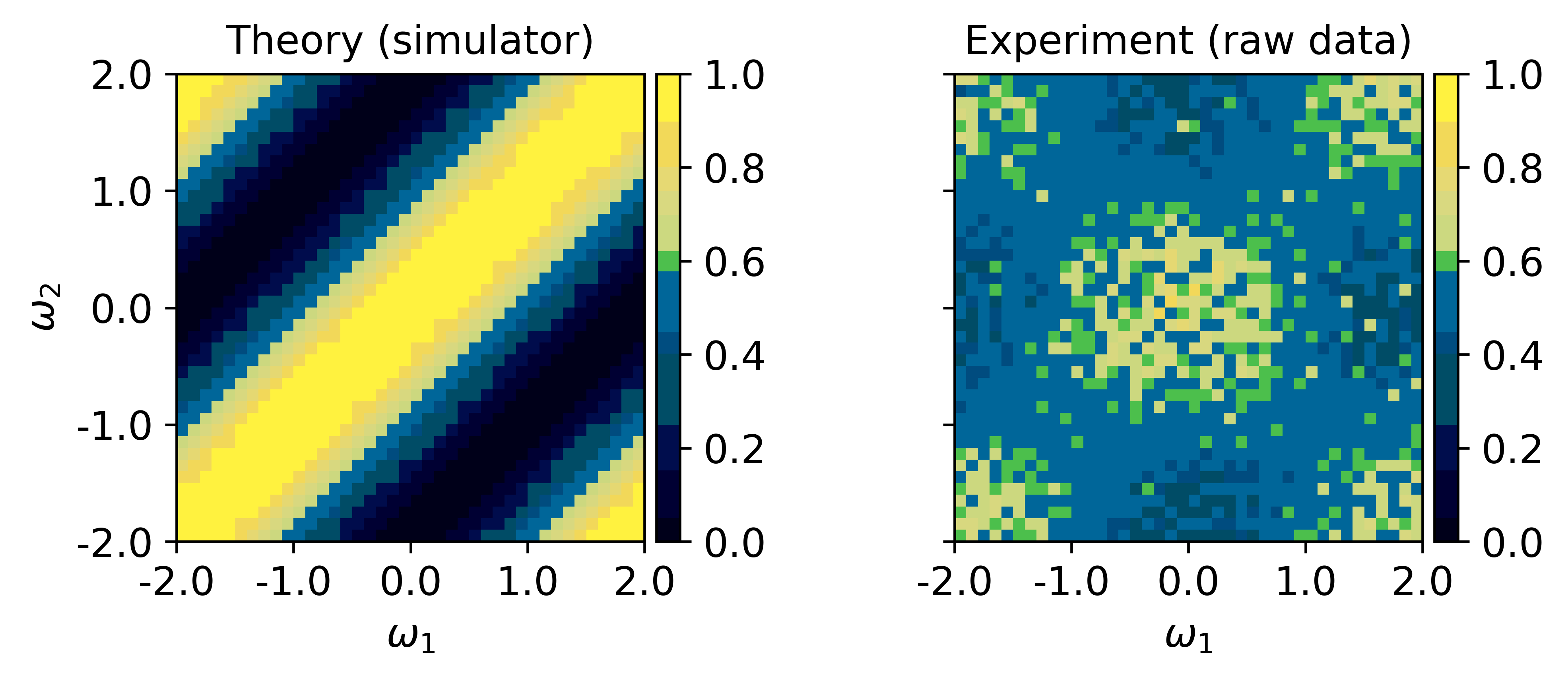}}
	\caption{$P_0(\ket{\Psi_{-}}; \omega_1,\omega_2)$ obtained from IBM classical simulator and the real quantum device.}	
\end{figure}

As a measure of difference between ideal (theoretical) results and the experimental data we have chosen several standard metrics:\\

\noindent (i) a signal-to-noise measure, defined as
\begin{equation}
SNR(M, M^{\prime}) = 10log_{10}\biggl(\frac{\sigma^{2}(M)}{MSE(M,M^{\prime})}\biggl),
\end{equation}
where $M$ and $M^{\prime}$ are arrays of data, obtained from the quantum chip and ideal classical simulator, correspondingly, $\sigma^{2}(M)$ is variance and $MSE(M,M^{\prime})$ is mean square error. 

\noindent (ii) $L_{1}$ distance (Manhattan distance), defined as 
\begin{equation}
    d_{L_{1}}(M, M^{'}) = \sum_{m \in M, m^{'} \in M^{'}}|m - m^{'}|,
\end{equation}
where $m$ and $m^{'}$ are elements of matrices $M$ and $M^{'}$ correspondingly.

\noindent (iii) Pearson correlation, defined as
\begin{equation}
    \rho_{M, M^{'}} = \frac{E[(M - E[M])(M^{'} - E[M^{'}])]}{\sigma(M)\sigma(M^{'})},
\end{equation}
where $E[M]$ is an expectation value of $M$.

We are going to trace the evolution of these three quantities after each step of our denoising procedure.



\subsection{Postselection}

The first step of our procedure is associated with the postselection of experimental data. The underlying idea is the following: consider we run some quantum circuit on a noisy quantum device and there are certain constraints on possible outputs. These constraints can originate from, e.g., symmetric considerations and the knowledge of constraints does not necessary require the resolution of the full problem -- otherwise, quantum computer is useless. For example, in simulations of many-body systems there may be certain conditions dictated by an electron-hole symmetry or particle-number conservation. Thus, in computations with noisy quantum devices, we may discard wrong outputs which explicitly violate such requirements. Note that some of us have recently used this idea in Ref. \cite{weare1} dealing with benchmarking of quantum computers using quantum communication protocols.

In the situation we here consider similar constraints can be deduced from the explicit derivation of the circuit's output. Since the main goal of this part of our paper is linked to error mitigation, it is legitimate to use some information from the explicit treatment. Namely, under the proper work of the quantum machine, if the input state is $\ket{\Phi_{-}}$, the output must be a superposition of $\ket{\Phi_{-}}$ and $\ket{\Phi_{+}}$ irrespective of $(\omega_{1}$, $\omega_{2})$. Thus, if the result of measurements of two register qubits in the computational basis is $00$ or $11$, this result can be discarded. In order to perform such a postselection, we need to measure not only ancillas, but also data qubits.

The approach we use is not completely universal, since it relies on constraints or symmetries which do not exist for an arbitrary problem. However, we would like to stress that, under certain conditions, it may be efficient to use a redundant coding, i.e., to encode a single logical qubit into larger number of physical qubits. Automatic error correction or classical postselection of results can be then applied to discard part of wrong outputs associated with certain quantum errors. Of course, a redundant coding is associated with the increase of the number of noisy gates of the algorithm, but nevertheless the advantages due to the postselection can overcome disadvantages due to the increase of the gate number. The success of the this strategy depends on the details of the algorithm as well as on the errors mechanisms and errors rates. For example, in \cite{weare1} the redundant encoding supplemented by the postselection was utilized and certain improvement of results has been achieved.    

The results of a postselection for the problem we here address are shown in Fig. 7, while Table 1 provides metric values before the postselection and after it. All three quantities indicate certain improvement of data after the procedure we utilized. However, there are also some qualitative changes in the overall distribution of the probability, which can be noticed by comparing experimental data after postselection and raw experimental data (Fig. 7). Namely, postselection leads to the emergence of a correct paternal structure of probability distribution -- separate "islands" now tend to be connected by "bridges". This fact is crucial for the subsequent analysis, since it allows for the partial reconstruction of correct data at the end of our procedure.

The fraction of discarded data after this step is approximately $1/2$ and it is not so dependent on $(\omega_{1}$, $\omega_{2})$. Of course, the additional measurement of two qubits leads to the increase of total readout error rate. However, these extra errors are definitely much smaller than the total error accumulated by the whole algorithm. This conclusion is evident from the fraction of discarded results, which is as high as $1/2$, and known error rates of readouts the latter being typically only several percent.

\begin{figure}[h!]
	\center{\includegraphics[width=1.0\linewidth]{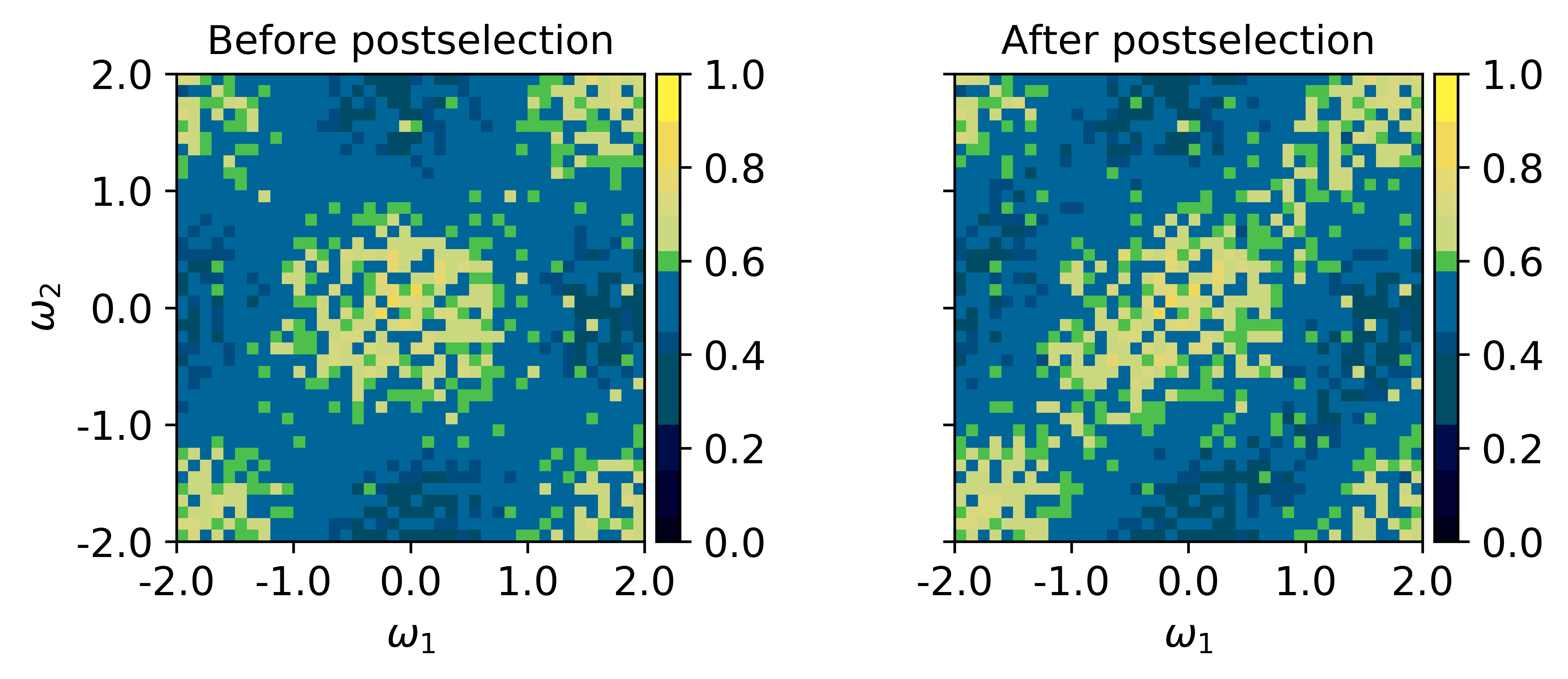}}
	\caption{$P_0(\ket{\Psi_{-}}; \omega_1,\omega_2)$ before and after the postselection procedure.}	
\end{figure}

\begin{center}
	\begin{tabular}[t]{|c|c|c|c|}
		\hline
		& No postselection & Postselection \\
		\hline
		SNR & -11.758 & -10.669\\
		\hline
		$L_{1}$ & 0.566 & 0.544\\
		\hline
		Pearson & 0.551 & 0.644\\
		\hline
		
	\end{tabular}\\
	\captionof{table}{Metrics values before the postselection and after it.}\label{sophisticatedtable}
\end{center}

\subsection{Image denoising}

Let us now discuss another series of heuristic tricks we use to partially suppress the effects of noises. They are associated with the image denoising. However, before that let us stress that postselection should be used before this step, otherwise the reconstruction will completely fail. Particularly, without the postselection, the probability pattern lacks connecting "bridges" between "islands" we mentioned before. These features are, of course, crucial for the reconstruction of a correct pattern.

We start with the observation that the experimentally determined values of probability are generally close to $0.5$ instead of being distributed between $0$ and $1$. Nevertheless, the spatial variations of probability as well as its pattern structure in $(\omega_1, \omega_2)$ plane are reproduced much more adequately. Notice that both controlling parameters $(\omega_1, \omega_2)$ enter the circuit only through single-qubit rotations. The obtained results imply that, in our noisy experiments with real hardware, the output results can be roughly divided into two classes: (i) wrong outputs, which are due to the single or multiple errors occurring during the algorithm executions and (ii) correct results corresponding to zero number of errors occurred. The first contribution is apparently dominant. An important observation is that it is nearly independent of controlling parameters $(\omega_1, \omega_2)$. A similar behavior has been recently observed by some of us in Ref. \cite{weare} dealing with the simulation of unitary evolution of spin clusters using programmable quantum hardware, where a similar controlling parameter was associated with the dimensionless time. The uniformity of wrong part of the output data with respect to this parameter was attributed to the fact that the circuit was not so shallow and contained a reasonable number of noisy quantum gates. An error occurring at particular gate produces its own dependence of the corresponding output on the controlling parameter. However, such dependencies for errors occurring at different gates of the circuit are also different, so that they finally average out into a nearly uniform dependence on the controlling parameter. Hence, this nearly uniform "background" can be simply eliminated by considering properly normalized differences instead of absolute values of quantities of interest. Let us stress that this situation is a direct consequence of a relatively large number of noisy gates in the circuit -- noise in this regime, in some sense, can help extracting valuable information from imperfect data. Of course, as the number of noisy gates grows, the fraction of correct outputs lowers down exponentially -- as a result, the trick we discuss can be utilized only in the regime of "intermediate-depth" circuits.

In order to get rid of background, we apply the following transform:
\begin{equation}
P'_0 = \frac{P_0- \displaystyle{\min P_0}}{\displaystyle{\max P_0} - \displaystyle{\min P_0 }},
\label{matrix}
\end{equation}
where we introduced the notation $P_0=P_0(\ket{\Psi_{-}}; \omega_1,\omega_2)$.
This transform rescales linearly the measured quantity in such a way that the lowest value is mapped to 0 and the highest value is brought to 1. We point out that this trick is not a fitting to the already known result. Our methodology is that, in our reconstruction, we use only a partial information on a correct and unknown probability distribution, which in this case is just the minimum and maximum value of the quantity of interest. In many cases, such additional parameters can be deduced from quite general considerations and do not require full knowledge of the output from the quantum computer. The result of the procedure is shown in Fig. 8, and Table 2 gives an evolution of metric values. We see that SNR was improved as well as the Manhattan distance. However, this is not true for the Pearson coefficient which did not change. This latter result is natural, since the Pearson coefficient must be insensitive to linear transformations.

Although the transformation defined by Eq. (\ref{matrix}) enables us to partially get rid of the nearly constant background, it has a serious drawback. The problem is that only a single value of probability corresponding to some particular point of the map is brought to 1, while the probability generally fluctuates significantly from one discrete point in $(\omega_1, \omega_2)$ plane to another. The origin of these fluctuations is associated with imperfections of quantum gates.

The particular point of maximum probability resides nearly at the center of the map shown in Fig. 6, i.e., at $\omega_1, \omega_2 \approx 0$. The same problem, of course, exists for the particular point of the map, for which the measured probability is lowest and hence is switched to 0 by rescaling (7). In order to circumvent this problem we apply a well known sigmoid transformation. It maps  $P'_0$ to the new value $P''_0$, according to
		\begin{equation}
		P''_0 = \frac{1}{1 + exp(-a(P'_0 - b))},
		\end{equation}
		where $a$ and $b$ are free parameters. The value of $b$ is fixed by the requirement that $b$ must stay invariant under the transformation, so that $b=0.5$ in our case.

Again, from general considerations, we can deduce a partial information about a true probability pattern, which includes not only minimum and maximum values of this quantity, but also a typical length scale of its variation in the space of parameters $(\omega_1, \omega_2)$. For the $2-$qubit input state and the problem we here consider this length scale can be roughly estimated as $\approx 1/2$. Next, we can define another length scale which is much smaller and evaluate the mean value of probability over the corresponding area. It is clear that the probability must be essentially constant within this area. Thus, we choose the parameters of the sigmoid transformation $a$ in such a way as to map the mean value of probability within the corresponding area $<f>_{max}$ in the vicinity of its maximum to some number, which is slightly lower than 1 (or alternatively, slightly higher than 0 in the vicinity of its minimum). We choose this number as $0.9$. This leads us to equate the $P''_0 (<f>_{max})$ and $0.9$.  We thus find $a \approx 5/(2<f>_{max}-1)$. We obtained that $<f>_{max}$  for our set of data is nearly 0.65 in the close vicinity of the point $\omega_1, \omega_2 \approx 0$ (averaging has been performed over the area of $5 \times 5$ points) and hence $a\approx 15$. Let us stress that the quality of reconstruction is nearly the same until $a$ ranges from 10 to 20, thus the choice of a characteristic number 0.9 as well as the area of the region for performing averaging are rather relative.

Table 2 provides the evolution of results for the metrics values. The use of the sigmoid transformation with $a=15$ applied after the normalization gives further improvement of data quality according to the SNR metrics. However, $L_1$ and Pearson coefficients indicate certain decrease of the agreement between the experiment and theory. The reason is linked to the fact that the sigmoid transformation, at this stage, produces artifacts -- it enhances fluctuations in some points of the plane by bringing values of probabilities close either to 0 or to 1. It is evident that $L_1$ is a point-wise local metric and it is rather sensitive to the enhancement of such local fluctuations. Pearson coefficient is also more sensitive to local fluctuation than SNR which is consistent with the fact that it is invariant under the rescaling of the probability pattern as a whole.  

\begin{center}
	\begin{tabular}[t]{|c|c|c|c|c|c|c|}
		\hline
		Step number & 0 & 1 & 2 & 3 & 4 & 5 \\
		\hline
		SNR & -10.669 & -6.016 & -0.520 & -9.457 & 1.406 & 7.387 \\
		\hline
		$L_{1}$ & 0.544 & 0.463 & 0.531 & 0.467 & 0.320 & 0.229 \\
		\hline
		Pearson & 0.644 & 0.644 & 0.616 & 0.896 & 0.896 & 0.911 \\
		\hline
	\end{tabular}\\
	\captionof{table}{Metrics values evolution during the following postprocessing procedure: postselection (step 0) $\rightarrow$ normalization (step 1) $\rightarrow$ sigmoid transform (step 2) $\rightarrow$ mean filtering (step 3) $\rightarrow$ normalization (step 4) $\rightarrow$ sigmoid transform (step 5)}\label{sophisticatedtable}
\end{center}

The procedures we have used do not completely suppress fluctuations of probability between neighboring discrete points of the map, moreover, the sigmoid transformation even enhances them to a certain extent. A natural idea is to use a mean filtering, i.e., to average out discrete data over small areas discussed in relation to the sigmoid transformation. However, this leads to the fact that the probability is again shifted towards $0.5$. As seen from the results of Table 2, it is accompanied by the decrease of SNR, although other metrics show better results due to the fact, that after filtering procedure the artifacts of sigmoid transformation, as discussed above, have been partially suppressed. In order to get rid of the decrease of SNR at this stage, we afterwards re-apply normalization and sigmoid transforms with the same parameters $a$, $b$ and achieve a further improvement of data quality according to the three metrics we used.

The whole procedure of postprocessing is the following: postselection (step 0) $\rightarrow$ normalization (step 1) $\rightarrow$ sigmoid transform (step 2) $\rightarrow$ mean filtering (step 3) $\rightarrow$ normalization (step 4) $\rightarrow$ sigmoid transform (step 5). The final result at the end of last three steps of this sequence are shown in Fig. 8. Table 2 provides an evolution of metrics values, which shows that all of them have been significantly improved although the details of their evolution at different steps of our procedure were not identical due to different types of correlations these quantities are responsible for. The comparison between the final pattern and the exact pattern shows that the agreement is good, although certain discrepancies are still present. As a whole, the improvement compared to raw data is significant. Thus, our procedure provides a case study which illustrates that it is possible to extract valuable information from data of noisy quantum computer even if they are heavily damaged by the decoherence and gate errors.

\begin{figure}[h!]
\center{\includegraphics[width=1.0\linewidth]{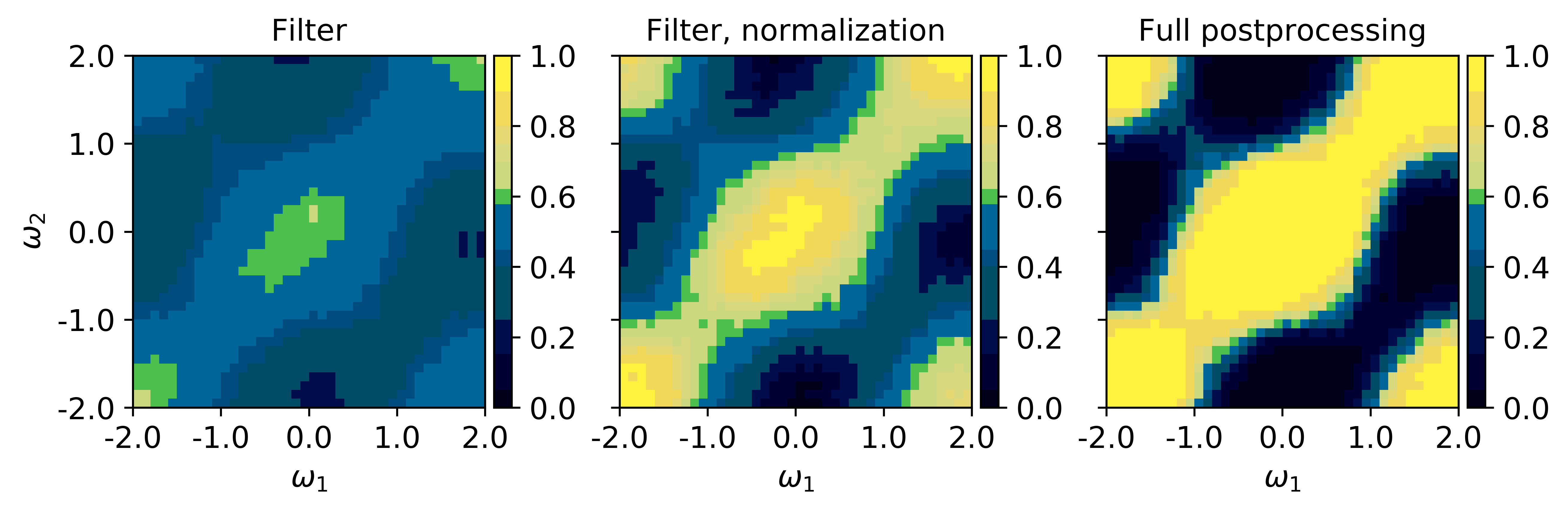}}
	\caption{Probability patterns after three last steps of the postprocessing procedure (see in the text).}	
\end{figure}

\section{Conclusion}

In this paper, we have addressed a hybrid quantum-classical scheme for the classification of input quantum states, where quantum part is represented by the phase estimation algorithm. It is based on a tunable unitary operator which can be adjusted to accomplish a desired classification of input quantum states from the training set. Due to the fact that measurements are performed on ancilla qubits, the classification can be made nondestructive and deterministic. For a general input quantum state, the scheme works as a probabilistic classifier and can be used to classify underlying patterns in quantum data.

We demonstrated proof-of-principle implementation of this idea using a superconducting quantum computer of IBM Quantum Experience and a specific simple example of the hybrid scheme we suggested. This scheme is able to classify maximally entangled two-qubit states into two groups depending on their parity. The real quantum hardware is characterized by different imperfections which lead to the accumulation of errors during the algorithm executions. Error mitigation, within our realization, was another issue addressed in this paper. We have applied a series of tricks associated with classical postprocessing to improve the raw experimental data and to recognize patterns contained in them. These ideas may be used in other realizations of hybrid quantum-classical computation schemes. Our results also demonstrate that pattern recognition can be an important ingredient of classical postprocessing of data from noisy quantum hardware.

\textbf{Acknowledgments.} -- We acknowledge use of the IBM Quantum Experience for this
    work. The viewpoints expressed are those of the authors and
    do not reflect the official policy or position of IBM or the
    IBM Quantum Experience team. W. V. P. acknowledges a support from RFBR (project no. 19-02-00421).

\end{document}